\documentclass[english]{aastex}
\usepackage[T1]{fontenc}
\usepackage{array}
\usepackage{textcomp}
\usepackage{multirow}
\usepackage{amstext}
\usepackage{graphicx}
\usepackage{subscript}

\makeatletter

\newcommand{\noun}[1]{\textsc{#1}}
\DeclareRobustCommand{\greektext}{%
  \fontencoding{LGR}\selectfont\def\encodingdefault{LGR}}
\DeclareRobustCommand{\textgreek}[1]{\leavevmode{\greektext #1}}
\DeclareFontEncoding{LGR}{}{}
\DeclareTextSymbol{\~}{LGR}{126}
\providecommand{\tabularnewline}{\\}

\slugcomment{}
\shorttitle{}
\shortauthors{}

\usepackage{babel}

\renewcommand\[{\begin{equation}}
\renewcommand\]{\end{equation}}

\makeatother

\usepackage{babel}
\begin{document}

\title{The Variable Optical Polarization and \emph{FERMI} Observations of
PMN J0948+0022}

\author{Joseph R. Eggen, H. Richard Miller, Jeremy D. Maune}

\affil{Department of Physics and Astronomy, Georgia State University, Atlanta,
GA 30303-3083}

\email{eggen@chara.gsu.edu}
\begin{abstract}
We report on observations of the \textgreek{g}-ray and optical photopolarimetric
behavior of the radio-loud, narrow line type-1 Seyfert galaxy PMN
J0948+0022 over a twenty seven month period. As this object has recently
been suggested to represent a prototype of an emerging class of blazar-like
objects, the observed properties are compared to those of blazars.
We extract doubling timescales of roughly 4 hours for the optical
and \textgreek{g}-ray bands. The rapid microvariability in the optical/NIR,
significant and variable optical polarization, and strong yet rapidly-variable
\textgreek{g}-ray emission we observe for PMN J0948+0022 are all classical
observational characteristcs associated with blazars. However, since
these observations do not show a clear correlation between the \textgreek{g}-ray
and optical behavior, they do not offer conclusive proof that the
emissive behavior of PMN J0948+0022 is due to a relativistic jet oriented
close to our line of sight. 
\end{abstract}


\section{Introduction}

Recently, several members of a sub-class of active galactic nuclei
have been observed with properties that would have previously been
divided among Seyfert galaxies, Broad Line Radio Galaxies (BLRG),
and blazars. These objects, the Radio-Loud Narrow Line Type-1 Seyferts
(RL NLS1), possess the standard identifying properties of NLS1 (\citealp{Osterbrock & Pogge,Kellermann et al 1989}):
strong optical emission of FeII, weak emission from forbidden lines
(i.e. {[}OIII{]}/H\textgreek{b} < 3), and FWHM(H\textgreek{b}) \ensuremath{\le}
2000 km/s \citep{Goodrich1}. However, the property of radio-loudness
(R \ensuremath{\ge} 10, where R = f\textsubscript{5.0 GHz}/f\textsubscript{4400\AA{}})
\citep{Kellermann et al 1989} is markedly rare in galaxies of this
type, occurring in < 7\% of such systems \citep{Komossa et al 2006}.
High brightness temperature ($\geq10^{13}$K), radio \citep{Zhou et al 2003}
loudness, and strong/rapid variability are, however, properties of
blazars. It is the combination of these various observational properties
that lead many to now believe that RL NLS1 and blazars may both play
host to relativistic jets.

It is now widely accepted that all varieties of Active Galactic Nuclei
(AGN) are manifestations of the same basic phenomenon - accretion
of matter onto a Supermassive Black Hole (SMBH) at the center of a
galaxy. The different classes of AGN that we then observe result,
to a large degree, from these objects being oriented differently with
respect to our line of sight. Blazars, a class of AGN characterized
by strong and variable emission across all wavelengths and strong
and highly variable polarization in the radio and optical \citep{Blandford & Rees},
are believed to result from the orientation of our line-of-sight near
the axis of a relativistic jet of particles being emitted from the
central engine of the source. Until recently, blazers were almost
exclusively observed to be hosted in elliptical galaxies, with very
few exceptions \citep{McHardy et al 1994}. Since the majority of
NLS1 hosts are spiral galaxies, finding evidence of blazar-like behavior
in such systems would help to fill in a curiously barren demographic
of the blazar population.

PMN J0948+0022 is an object that displays the expected properties
of a Narrow-Line Seyfert-1 galaxy as described above \citep{Zhou et al 2003},
as well as those of blazars, such as strong and variable emission
in the radio thru \textgreek{g}-ray energies over long timescales
\citep{Abdo et al 2009,Foschini et al 2012}, and microvariability
in the optical \citep{Maune et al 2013}. \citet{Ikejiri et al 2011}
also observed PMN J0948+0022 to exhibit a very high degree of linear
polarization (18.8\%) in the optical (V-band), when the object was
very bright (V = 17.028 $\pm$ 0.014). 

To date, no comprehensive, long-term program investigating the optical
polarimetric/photometric characteristics of RL NLS1 has been reported.
This paper provides the results of such a study for the prototype
for this class of objects, PMN J0948+0022.

Throughout this manuscript, Julian Dates (JD) are expressed as Modified
Julian Dates MJD. The conversion to MJD is expressed as MJD = JD -
2.45e6.

\section{Observations and Data Reductions}

\subsection{Optical Photopolarimetry Data}

All optical polarimetric data used in this study were obtained with
the 72-inch Perkins telescope at Lowell Observatory in Flagstaff,
Arizona, using the PRISM instrument which includes a polarimeter with
a rotating half-wave plate. Data were obtained during several observing
runs between February, 2011 and April, 2013. The specific dates of
each observation are given in Table 1. The observations consisted
of a series of 2-4 measurements for the Q and U Stokes parameters
per polarization observation. Each series consisted of four images,
each taken at different instrumental position angles \textendash{}
0\textdegree{}, 45\textdegree{}, 90\textdegree{}, and 135\textdegree{}
\textendash{} of the waveplate. 

Corrections to polarimetric values were obtained from two sources:
in-field comparison stars and seperately-observed polarimetric standards,
both polarized and unpolarized \citep{Schmidt et al 1992}. As the
camera has a wide field of view (approx. 14' x 14'), we are able to
use field stars for interstellar polarization corrections by subtracting
the average percent polarization of the brightest field stars. Polarized
and unpolarized standard stars are used to calibrate corrections for
polarization Position Angle (P.A.) and instrumental polarization (typically
less than 1\%) \citep{Jorstad et al 2010}, respectively.

The data were reduced and analyzed using in-house scripts, which utilize
standard packages in the PyRAF 2.0 suite of reduction tools%
\footnote{PyRAF is a product of the Space Telescope Science Institute, which
is operated by AURA for NASA.%
}. Bias frames were taken at the beginning of every night and combined
into a master bias that was subtracted from each image. Flat frames
were taken at least once per run, using a featureless screen inside
the dome. Each position of the waveplate required its own set of flats,
which would later be combined into one master flat per position angle
for application to the appropriate science image(s). Cosmic ray cleaning
was performed on all science images, with the \textbf{threshold} and
\textbf{fluxratio} parameters set to 35 and 5, respectively. Aperture
photometry was then performed on the calibrated science frames on
an object-by-object basis. An aperture radius of 7 arcsec was used
on all images both to maximize the signal-to-noise and to maintain
consistency with the optical photometry being performed on this target
by \citet{Maune et al 2013}. Use of the in-field comparison stars
compiled by Maune et al. allowed for one to obtain simultaneous measures
of the absolute R-band magnitude, as well as the percent polarization
(P) and electric vector position angle (EVPA).

For ease-of-comparison with \textgreek{g}-ray data, which are expressed
as photon fluxes in this manuscript, optical data were converted from
magnitudes to units of flux (mJy) using the following equation:

\[
F=2941*10^{-0.4*Mag}
\]
\label{(1)}Where \emph{F} is the flux in mJy and \emph{Mag} is the
R-band magnitude.

\subsection{Optical and Infrared Photometry with SMARTS}

Much of our optical and all of our NIR data were obtained by the 1.3m
telescope at the Cerro Tololo Inter-American Observatory (CTIO) under
the Small and Moderate Aperture Research Telescope System (SMARTS)
program. We obtained simultaneous data in the optical R and infrared
J bands using ANDICAM, which is a dual-channel instrument that uses
a dichroic to simultaneously feed optical and IR CCD imagers, allowing
the acquisition of IR data from 0.4 to 2.2 \textgreek{m}m. Our limited
dataset of J-band images consisted of four NIR images - one for each
corresponding optical R-band image - which were flat-fielded, overscan-corrected,
bias-subtracted, and co-added using standard PyRAF/IRAF packages and
scripts. To be consistent with our optical data, a 7 arcsec aperture
radius was used to perform differential photometry.

\subsection{\emph{FERMI}-LAT Data}

Gamma-ray data were obtained through the FERMI-LAT public data server.
The Large Area Telescope (LAT), on board the FERMI Gamma-ray Space
Telescope, is a pair-conversion detector sensitive to \textgreek{g}-rays
in the 20 MeV to several hundred GeV energy range \citep{Atwood et al 2009}.
The instrument has worked almost continuously in all-sky-survey mode
since its launch in June 2008, which allows coverage of the entire
\textgreek{g}-ray sky approximately every 3 hours. The data were reduced
and analyzed using ScienceTools v9r27p1 and instrument response functions
P7SOURCE\_V6. We utilized the likelihood analysis procedure as described
at the FSSC website. Photon fluxes were calculated using data from
MJD 5562 to MJD 6408 (January 01, 2011 to April 25, 2013).

Our data were downloaded from the FERMI website on March 27, 2013
and cover a region on the sky 15\textdegree{} in radius, centered
on the location of PMN J0948+0022 (2FGL0948.8+0020 from the Fermi
2-Year Point Source Catalog), and in an energy range of 100 MeV to
300 GeV. Our \textgreek{g}-ray light curve consists of 112 equally-sized
bins, each of which is 637861 seconds in length, or one-quarter of
the lunar synodic period, as our observing runs at Lowell Observatory
were centered around the time of the New Moon. The first bin began
on February 03, 2011, while the last bin ended on April 25, 2013.
Only data corresponding to the \noun{source} class (evclass=2) were
utilized, with a 52\textdegree{} cut-off rock-angle of the spacecraft,
while an additional cut utilizing an angle of 100\textdegree{} from
the zenith was imposed so as to minimize the contamination due to
\textgreek{g}-rays coming from Earth's upper atmosphere. Since PMN
J0948+0022 is within 20\textdegree{} of the ecliptic, and the Sun
is a source of \textgreek{g}-rays comparable to our target \citep{Abdo et al 2011b},
a final cut was used to exclude exposures that occurred when the Sun
was within the 10\textdegree{} Region of Interest (RoI). Photon fluxes
and spectral fits were derived using an unbinned maximum likelihood
analysis which was accomplished using the ScienceTool \noun{gtlike}.

In order to accurately measure the flux and spectral parameters of
the source, one needs to account for \textgreek{g}-rays emitted from
the background. To this end, two models were used: an isotropic background
model accounting for extragalactic diffuse emission and residual charged
particle background, and a Galactic diffuse emission model to account
for diffuse sources from within our own galaxy. The isotropic model
we used was the one contained in the file iso\_p7v6source.txt%
\footnote{http://fermi.gsfc.nasa.gov/ssc/data/access/lat/Background/Models.html%
}, while the Galactic component was given by the file gal\_2year7v6\_v0.fits.
The normalizations of both components were left to vary freely during
likelihood analysis.

In order to determine the significance of the \textgreek{g}-ray signal
from PMN J0948+0022, we used the Test Statistic (TS). The Test Statistic
is defined as TS = 2\textgreek{D}log(likelihood), where \emph{likelihood}
refers to the likelihood ratio test as described in \citet{Mattox et al 1996}.
Determining the likelihood of a given photon flux being produced by
a source with a given spectral model was accomplished using the \noun{gtlike}
Science Tool. Our source model consisted of all the known \textgreek{g}-ray
point sources located within a 15\textdegree{} radius of 2FGLJ0948.8+0020.
Initial values for all spectral parameters for these sources were
taken from the LAT 2-year Point Source Catalog. Along with PMN J0948+0022
and the aforementioned background models, several point sources were
allowed to vary (i.e. photon indices, normalization factors, and spectral
slope indices were left as free parameters) during the likelihood
analysis, so as to account for the inherent variability of many \textgreek{g}-ray
sources. The type of spectral model used for a given source was the
same model used for that source in the LAT 2-year catalog \citep{2FGL catalog}. 

For PMN J0948+0022 specifically, we used a LogParabola model to describe
the \textgreek{g}-ray spectrum of the source in this study. This model
takes the form:

\[
N(E)=N_{0}(\frac{E}{E_{b}})^{\text{\textgreek{a}+\textgreek{b}ln(E/\ensuremath{E_{b}})}}
\]
\label{(2)}where N\textsubscript{0} is the normalization index,
\textgreek{a} is the photon index at the pivot energy E\textsubscript{b},
and \textgreek{b} is the curvature index. These parameters were left
un-fixed with the exception of E\textsubscript{b}: this was fixed
at a value of 271.597 MeV, which is the value reported for this object
in the LAT 2-year catalog. All \textgreek{g}-ray fluxes presented
in this paper are integrated over the entire energy range cited above
(100 MeV to 300 GeV), with corresponding units of {[}ph cm\textsuperscript{-2}sec\textsuperscript{-1}{]}.

Sources outside our 10\textdegree{} Radius of Interest (RoI) but within
20\textdegree{} of the target were also included in the source model,
as the point spread functions of these objects could result in extra
photons seeping into the RoI of our target. All parameters for these
sources were fixed to their 2FGL catalog values during the analysis.

\section{Results}

\subsection{Polarimetry}

\begin{figure*}[t]
\noindent \centering{}\includegraphics[scale=0.75]{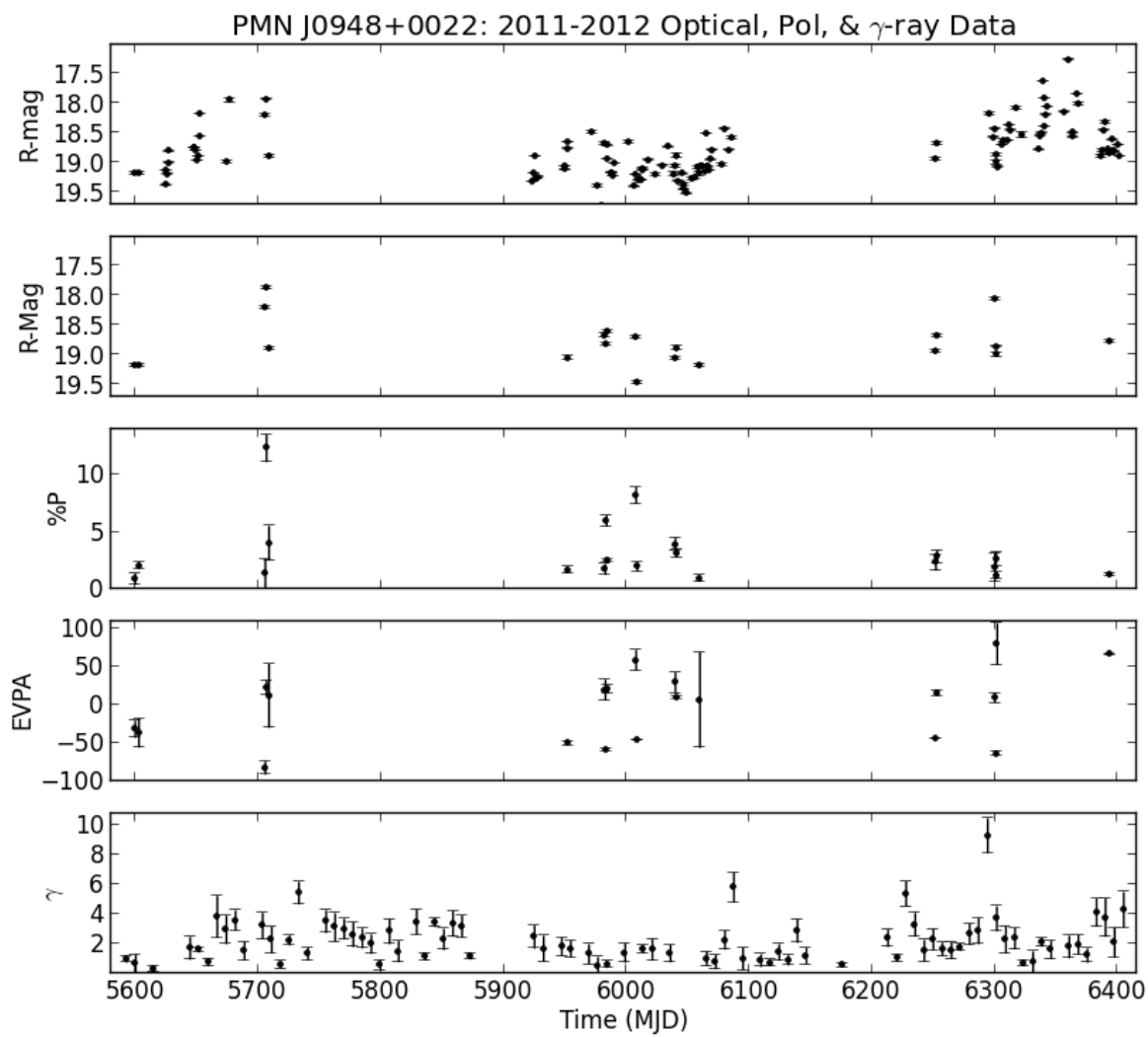}\caption{A comparison of all R-band optical data binned at 24-hour intervals
(top panel), only optical data obtained with polarimetry (second panel),
the Percent Polarization (third panel), position of the electric vector
in degrees (fourth panel), and the integrated \textgreek{g}-ray flux
(bottom panel) of PMN J0948+0022. Upper limits have been removed for
clarity. Details on the photopolarimetric data can be found in Table
\ref{tab:PhotPol-obs}. The same horizontal axis is common to all
five plots.\label{fig:opt_pol_gamma}}
\end{figure*}

Our polarimetric results are detailed in Table \ref{tab:PhotPol-obs}
and displayed for comparison in Fig. \ref{fig:opt_pol_gamma}. The
top panel of Figure \ref{fig:opt_pol_gamma} also displays the all
of our optical data for comparison, binned in 24-hour increments as
described in Section 3.2, below. Note that all photometric data points
in the aforementioned table and 2nd panel of Figure \ref{fig:opt_pol_gamma}
were derived from polarimetric measurements. PMN J0948+0022 displays
a moderate, but significant variability in both the percent polarization
(P) and EVPA, with a maximum P of 12.31 $\pm$ 1.21\% and an EVPA
which varied substantially. While there does seem to be evidence for
a correlation between the optical state and the polarimetric quantities
in the data - a high optical state coupled with a high value of P,
for example - the data also contains notable exceptions to such a
relationship (e.g., the lack of any increase in P during an outburst
in early 2013). No significant correlation was observed between the
optical or polarimetric quantities and the \textgreek{g}-ray flux.
It should also be noted that the 180\textdegree{} uncertainty inherent
in the measurement of the position angle may produce the appearance
of trends where none exist.

\begin{figure*}[t]
\noindent \centering{}\includegraphics[scale=0.75]{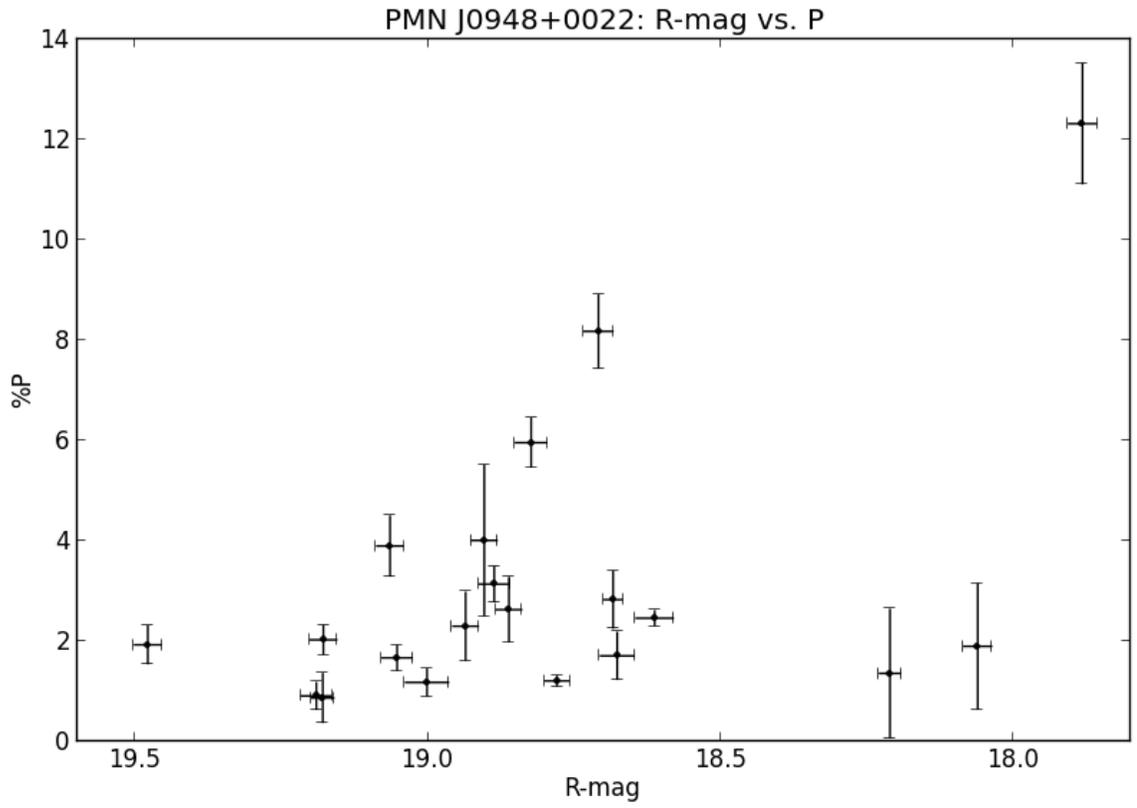}\caption{A plot of concurrent R-band magnitudes and P, based on the data in
Table \ref{tab:PhotPol-obs} so that each data point represents a
measurement of the value of P and the R-magnitude that coincide in
time. \label{fig:Rmag_vs_P}}
\end{figure*}

A plot of each value of P versus the concurrent R-magnitude is shown
in Figure \ref{fig:Rmag_vs_P}. A clear trend between these two values
is not immediately obvious. However, closer inspection of the three
brightest data points revealed notable characteristics. The brightest
data point (corresponding to the data from MJD = 5706.7 in Table \ref{tab:PhotPol-obs})
also represents the largest value of P and was taken when the object
was observed to be in a persistent (42-minute duration) and stable
(variation of 0.35 $\pm$ .05 magnitudes) bright state. The second-highest
point corresponds to the data at MJD = 6300.0 from Table \ref{tab:PhotPol-obs}
and occurred 4.44 hours after the object was observed to be 0.75 $\pm$
.02 magnitudes fainter via differential photometry. The third-brightest
point was obtained 24 hours before the brightest data point and though
they differ in time and brightness by relatively very little, they
have quite different polarimetric values. Possible interpretations
of these observations are discussed later in Section 4.

\subsection{R-band Photometry}

The optical photometric data utilized in the present study were obtained
from the following sources: (1) polarimetry obtained by the group
at Georgia State University (Table \ref{tab:PhotPol-obs}), (2) optical
data presented in \citet{Maune et al 2013}, and (3) additional data
collected since the publication of \citeauthor{Maune et al 2013}.
These datasets were merged to make a master optical light curve, which
contained 1321 R-band observations obtained between February 7, 2011,
and April 19, 2013. A program was written which binned the optical
data with the same temporal bounds as the \textgreek{g}-ray data.
An average value was then calculated for each bin to provide a single
data point (42 optical photometric data points in total, displayed
in Figure \ref{fig:IntGamma_with_opt}), allowing us to better-match
the optical photometric data to the cadence of the \emph{FERMI} observations.
We also binned the R-band data in 24-hour bins centered on 00:00 UT,
as the latitudes at which our data were obtained resulted in acquisition
times roughly centered on this time of day. The 24-hour binned data
(114 data points) can be seen in the top panel of Figure \ref{fig:opt_pol_gamma}
and are also accessible via the online version of this manuscript.
A sample of this data is provided in Table \ref{tab:1_day_bin_opt}.
This technique also served to ``smooth'' some of the large changes
in the optical flux over short time periods that are often manifest
in PMN J0948+0022 (see below).

By making use of high-cadence data presented in our previous work,
as well as new data obtained in January of 2013 , we were able to
make very precise measurements of the doubling/halving timescale (\textgreek{t})
for this object in the R-band. The formula for calculating the observed
timescale \textgreek{t} is given by the equation \citep{Foschini et al 2011}:

\[
F(t)=F(t_{0})*2{}^{-(t-t_{0})/\text{\textgreek{t}}}
\]
\label{(3)}where \emph{F(t)} and \emph{F(t\textsubscript{0}) }are
flux values at the times \emph{t} and \emph{t\textsubscript{0}},
respectively. Throughout the remainder of this manuscript we will
use \textgreek{t}\textsubscript{R} and \textgreek{t}\textsubscript{\textgreek{g}}
to refer to the doubling timescales measured for the optical R and
\textgreek{g}-ray bands, respectively.

\begin{figure*}[t]
\noindent \centering{}\includegraphics[angle=90,scale=0.75]{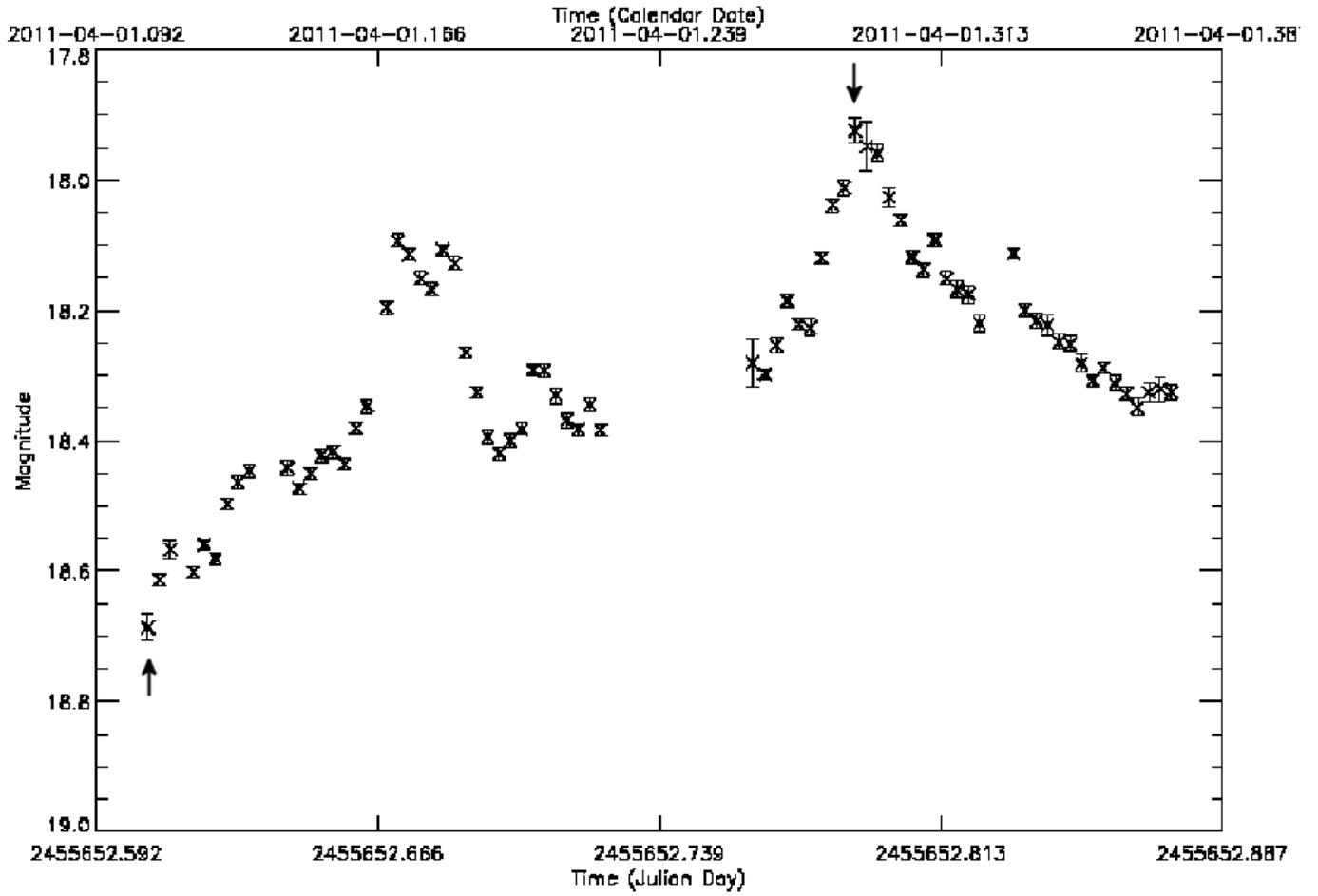}\caption{Microvariability data presented in \citet{Maune et al 2013}, showing
the doubling of the flux in 4.39 $\pm$ .19 hours. Arrows indicate
the data points separated by the necessary flux difference and used
to make the timescale calculation. Two discrete, roughly symmetric
flares can also be seen in this figure. \label{fig:doubling_timescale}}
\end{figure*}

\begin{figure*}[t]
\noindent \centering{}\includegraphics[angle=90,scale=0.75]{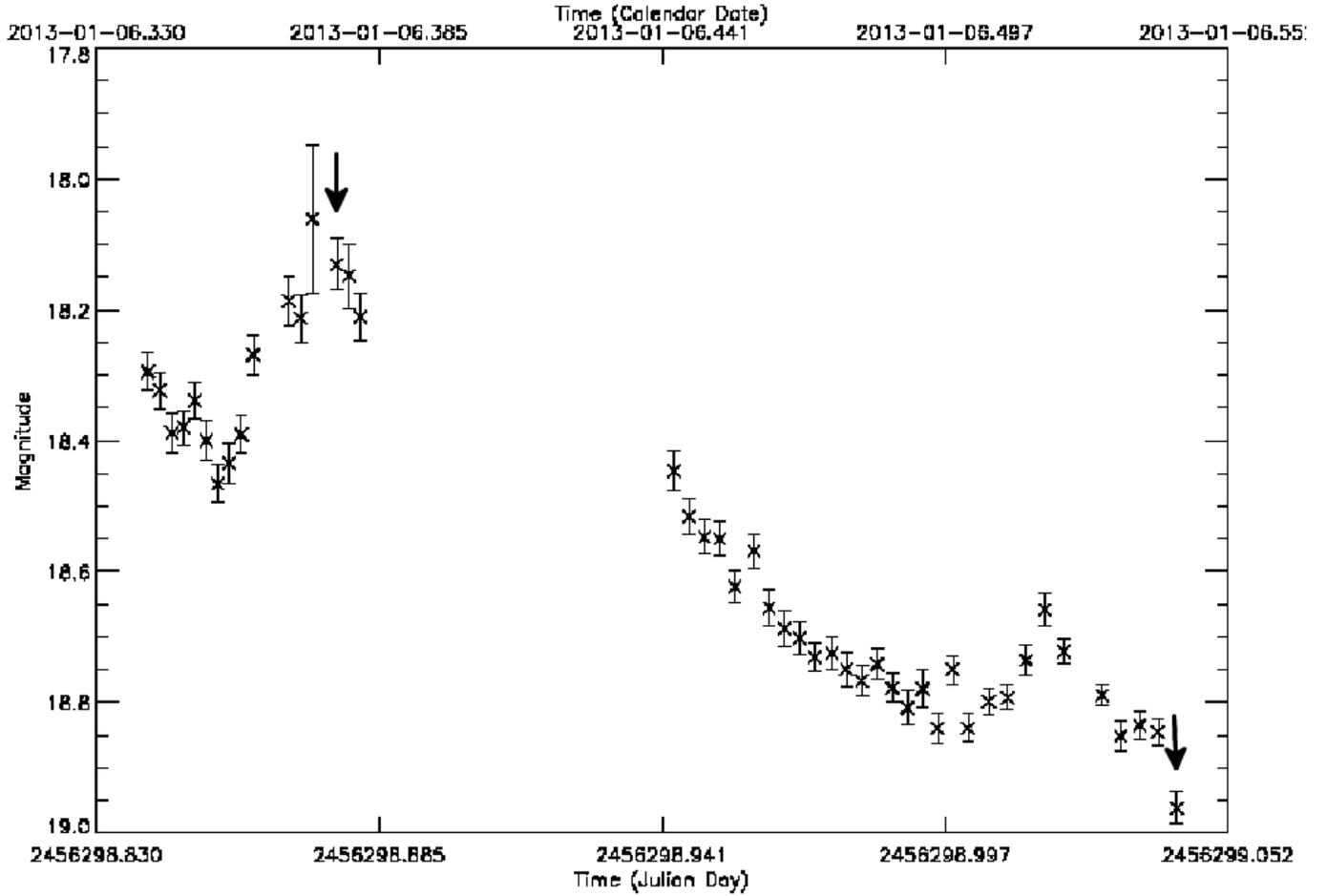}\caption{Similar to Figure \ref{fig:doubling_timescale}, but original to this
work and showing a halving of the total optical flux in 3.60 $\pm$
.23 hours. Note the discrete flare near the end of the lightcurve.\label{fig:halving_timescale}}
\end{figure*}

This quantity was calculated for PMN J0948+0022 in several wavebands
by \citet{Foschini et al 2012}, although they did not have access
to significant micro-variability data (especially in the optical bands)
in their study. Micro-variability data collected by the present authors
on two nights were used to calculate \textgreek{t}\textsubscript{R},
and are shown in Figures \ref{fig:doubling_timescale} and \ref{fig:halving_timescale}.
Figure \ref{fig:doubling_timescale} (first presented in \citet{Maune et al 2013})
reveals the target to be highly variable on very short (a few minutes)
timescales, with the object going from a minimum brightness of R =
18.69 $\pm$ .02 to a maximum of R = 17.92 $\pm$ .02 in 4.45 hours.
This resulted in a doubling timescale of \textgreek{t}\textsubscript{R}
= 4.39 $\pm$ .19 hours. Conversely, Figure \ref{fig:halving_timescale}
(original to this work) illustrates a rapid decrease in the brightness
of PMN J0948+0022, though over a similar magnitude range and duration
as the doubling event. Here we see the object fall from R = 18.13
$\pm$ .04 to R = 18.96 $\pm$ .02 in 3.97 hours, for a halving timescale
of \textgreek{t}\textsubscript{R} = 3.60 $\pm$ .23 hours. Together,
these lightcurves yield an average value of \textgreek{t}\textsubscript{R}
= 3.99 $\pm$ .15 hours. 

Concerning the aforementioned Figures, there are two items of note.
The first is that the above value for \textgreek{t}\textsubscript{R}
should be taken as an upper limit. The gaps in the lightcurves presented
in Figures \ref{fig:doubling_timescale} and \ref{fig:halving_timescale}
may contain higher or lower flux states which would drastically reduce
the doubling timescale. The second item is that both lightcurves show
discrete brightening events (flares - three in total) that are not
only well-sampled, but also roughly symmetric in their brightening
and dimming profiles and are on the order of an hour in duration.

\begin{figure*}[t]
\noindent \centering{}\includegraphics[angle=90,scale=0.75]{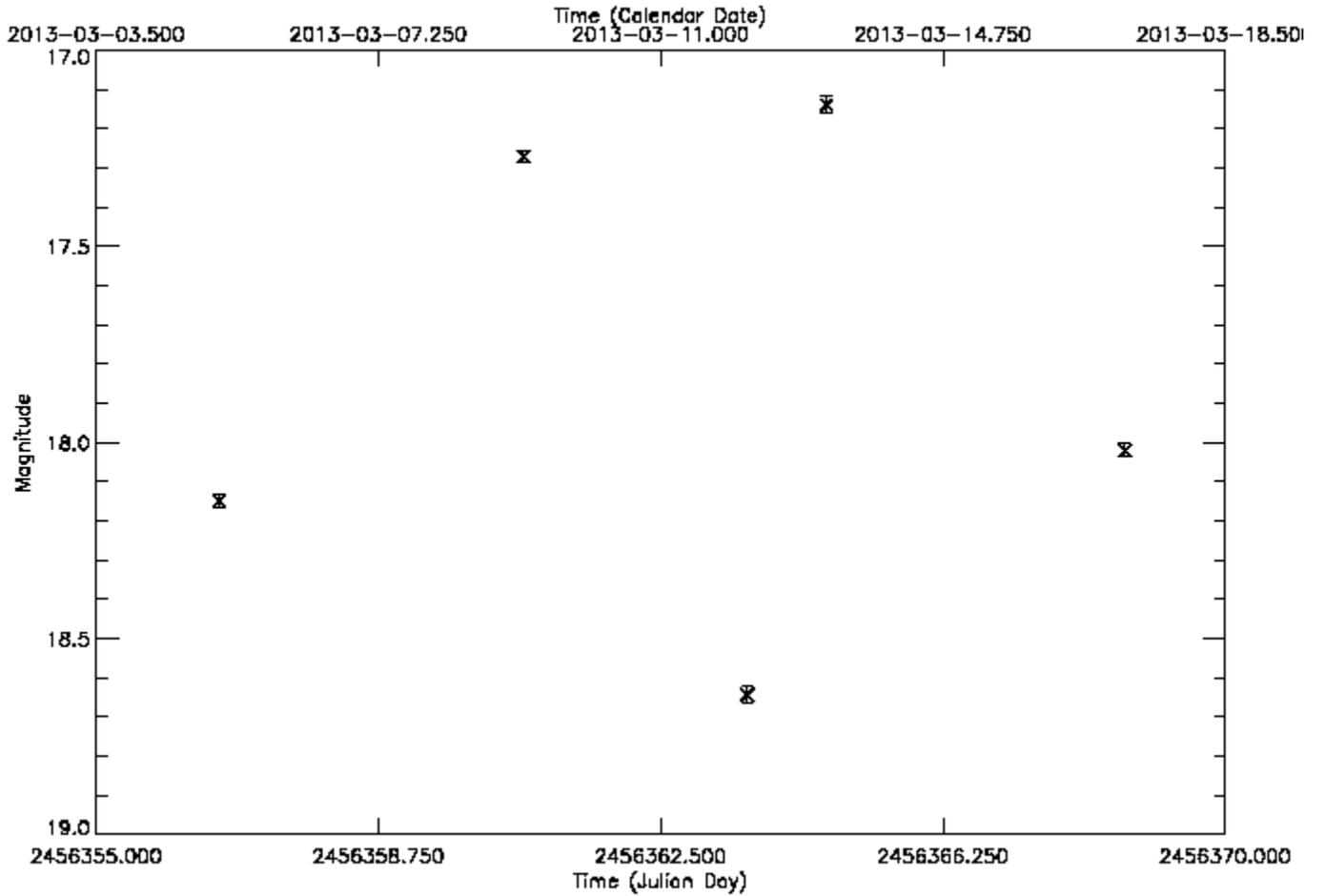}\caption{R-band optical data showing a rapid and dramatic rise in the flux
of PMN J0948+0022 which appeared to be uncorrelated with any \textgreek{g}-ray
activity. See Table \ref{tab:smarts_opt_ir} for the specifics of
the observations.\label{fig:orphan_flares}}
\end{figure*}

In March 2013, PMN J0948+0022 underwent a dramatic brightening in
the R-band (Figure \ref{fig:orphan_flares}), achieving a maximum
brightness of R = 17.140 $\pm$ 0.021, which was unprecedented for
the object in this band. We were fortunate to have obtained simultaneous
J-band data during this flaring episode, although we currently lack
calibrated values for field stars in the J-band. However, as can be
seen in Table \ref{tab:smarts_opt_ir}, the degree of the fractional
change of the optical flux was closely mirrored by that in the NIR
flux. Interestingly, there was no indication of correlated activity
in the \textgreek{g}-ray regime, on any timescale, which could be
linked to the optical/NIR activity. While correlation of activity
across multiple wavebands is a common characteristic of blazars (see,
for example, \citet{Rani et al 2013}, \citet{Abdo et al 2010}),
it is also not unusual to find examples of high activity in one band
without the presence of similar activity in other bands \citep{Krawczynski et al 2004}.

\subsection{\textgreek{g}-ray Observations}

\begin{figure*}[t]
\noindent \centering{}\includegraphics[scale=0.75]{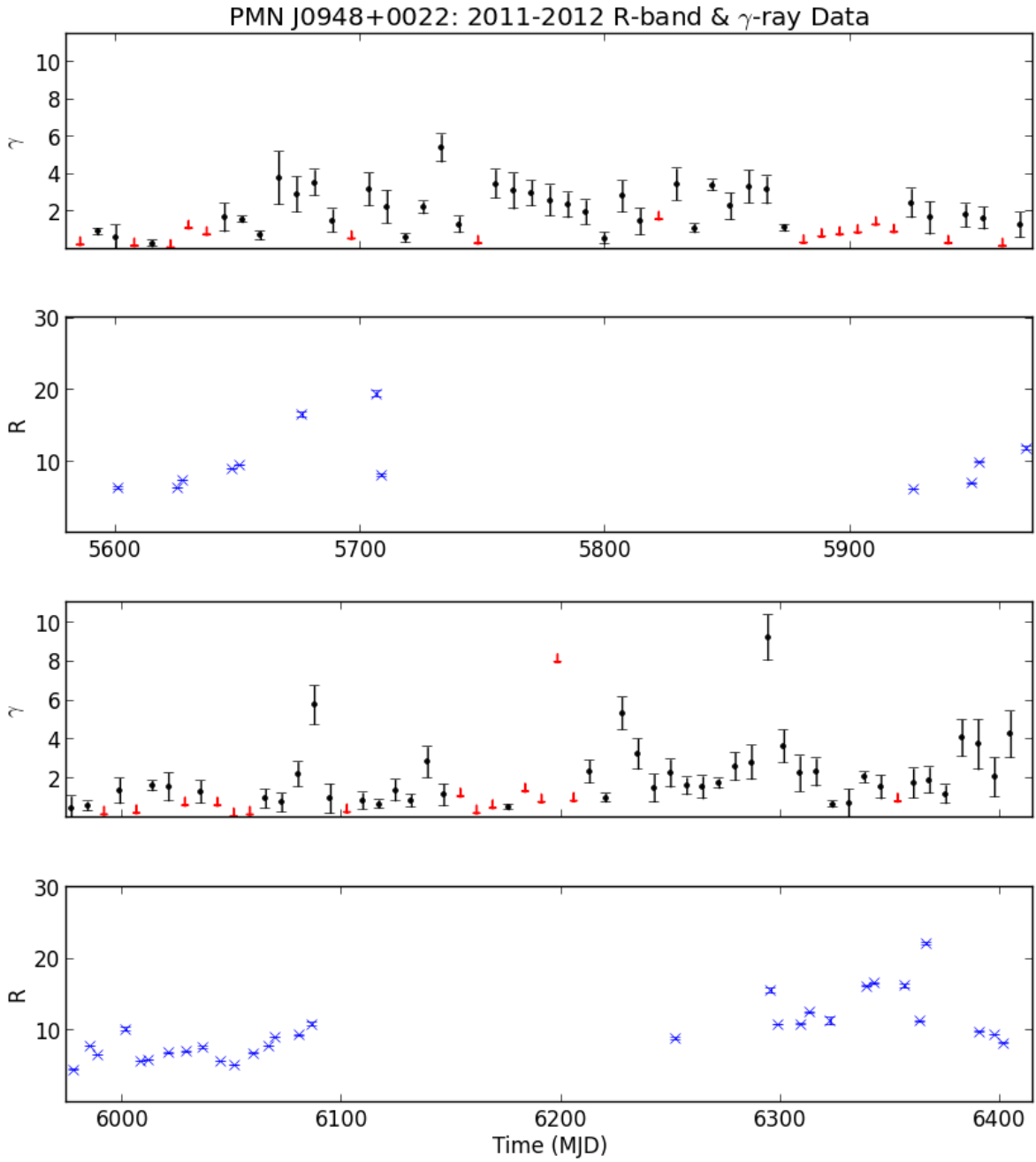}\caption{\textgreek{g}-ray flux (integrated from 100 MeV to 300 GeV) plotted
in panels 1 \& 3, with time-averaged R-band flux plotted in panels
2 \& 4. Panels 1 \& 2 share a common horizontal (time) axis, as do
panels 3 \& 4. Downward-pointing arrows denote upper-limits on \textgreek{g}-ray
data points. \label{fig:IntGamma_with_opt}}
\end{figure*}

The long-term behavior of PMN J0948+0022 in \textgreek{g}-rays is
shown in Fig. \ref{fig:IntGamma_with_opt}. Our long-term \textgreek{g}-ray
dataset consists of 112 data points in total, including 31 upper limit
values. The TS value may be used as a proxy for the confidence in
the measurement, with TS \ensuremath{\ge} 9 (corresponding roughly
to \textgreek{sv} \ensuremath{\ge} 3) \citep{Mattox et al 1996} being
the threshold above which we considered a positive detection. We applied
the method for extracting upper limits, as described on the \emph{FERMI}
website%
\footnote{http://fermi.gsfc.nasa.gov/ssc/data/analysis/scitools/python\_tutorial.html%
}, to bins with TS < 9. As PMN J0948+0022 is a well-established \textgreek{g}-ray
source \citep{Abdo et al 2009}, and less than 1/3 of our data points
are below this threshold, we feel it is reasonable to treat all of
our remaining gamma-ray data points as direct detections. 

PMN J0948+0022 was observed to undergo a significant increase in \textgreek{g}-ray
brightness in the spring of 2011, which was mirrored by a steady increase
of the average R-band luminosity over the next few months, as can
be seen in the top 2 panels of Fig. \ref{fig:IntGamma_with_opt}.
Unfortunately, the \textgreek{g}-ray brightening which was observed
in the spring of 2001 continued into the summer when PMN J0948+0022
could not be viewed in the optical due to the close proximity of the
Sun. Nevertheless, when optical monitoring resumed in November 2011,
the object was observed to be back at a low state in both the optical
and \textgreek{g}-rays and would remain in that state until the end
of optical monitoring in the spring of 2012. The resumption of optical
monitoring in the fall of 2012 was accompanied by an increase in \textgreek{g}-ray
activity which was dominated by two especially active periods centered
roughly on MJD 6225 and 6290. Perhaps most interestingly, the optical
activity remained generally much higher than average after MJD 6300,
even though \textgreek{g}-ray activity dropped to a relatively inactive
state.

\begin{figure*}[t]
\noindent \centering{}\includegraphics[scale=0.75]{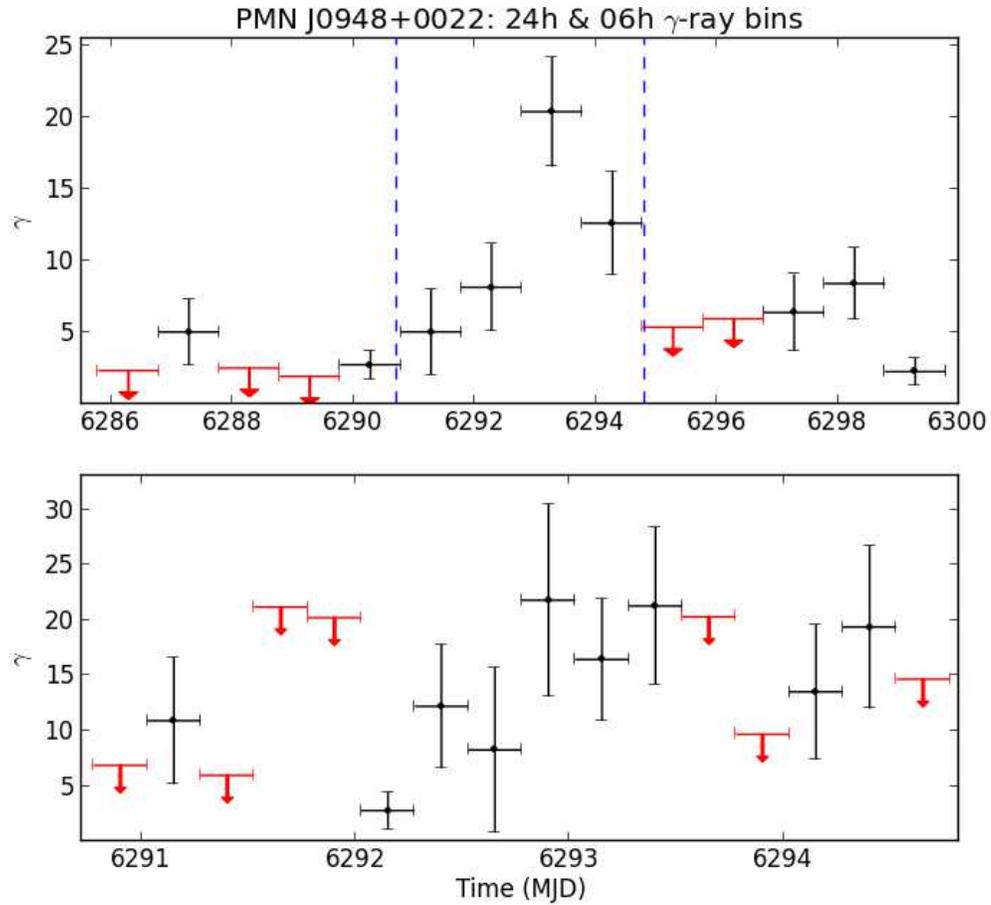}\caption{Top: \textgreek{g}-ray light curve centered on the high flux measurement
which occurred on MJD = 6294 in Figure \ref{fig:IntGamma_with_opt}
with 24-hour time bins. Bottom: The data circumscribed by the blue
dashed lines in the top panel, but analyzed with 6-hour time bins.\label{fig:zoom_24h_06h}}
\end{figure*}

On December 18, 2012, PMN J0948+0022 was observed to undergo a flare
in the near-infrared \citep{Carrasco et al 2012}, followed by a strong
\textgreek{g}-ray peak as observed by FERMI-LAT \citep{D'Ammando & Orienti 2013}.
The measured flux for the bin in which the peak of this event fell
(MJD = 6294) was 9.22x10\textsuperscript{-7} ph cm\textsuperscript{-2}s\textsuperscript{-1},
with a TS = 194.27. Due to the strength of this event, we were able
to bin the data to much higher cadence (6 hours) around the peak of
the flare. Figure \ref{fig:zoom_24h_06h} illustrates the rapidly-evolving
nature of the flare with 24-hour (top panel) and 6-hour (bottom panel)
bins. One quickly notices that the flux appears to double twice (within
a 2-\textgreek{sv} uncertainty) in the 6-hour light curve. Applying
Eq. \ref{(3)} to the relevant bins yields a measurement of the doubling
timescale in the \textgreek{g}-ray regime of \textgreek{t}\textsubscript{\textgreek{g}}
= 3.55 $\pm$ 4.29 hours - see Table \ref{tab:dh_timescale_stats}
for the values used to calculate this result. A doubling timescale
of approximately 4 hours not only roughly agrees with our measurement
of \textgreek{t}\textsubscript{R}, but is also comparable to the
\textgreek{g}-ray variability timescales found for 3C 454.3 during
that object's recent outbursts in 2009 \citep{Ackermann et al 2010}
and 2010 \citep{Abdo et al 2011a}, who found doubling timescales
of 3 hours and 6 hours, respectively.

\section{Discussion \& Conclusions}

Multi-epoch VLBA observations of a number of \textgreek{g}-ray detected
quasars and blazars \citep{Jorstad et al 2001} suggest that the \textgreek{g}-ray
emission observed for these objects originates near the radio core,
perhaps corresponding to a standing shock in the jet itself, and not
specifically originating from a location near the central SMBH. It
is thought that the oft-observed flares are produced as a result of
turbulence/instabilities, which are present in the relativistic jet
and arise as shocks in this outflow. With the passage of a shock down
the jet, the magnetic field will be compressed within the shock region
causing the field to become more highly ordered. Observationally,
one would then expect that the fractional linear polarization would
increase and the direction of the electric vector position angle would
change rapidly.

\begin{figure*}[t]
\noindent \centering{}\includegraphics[scale=0.75]{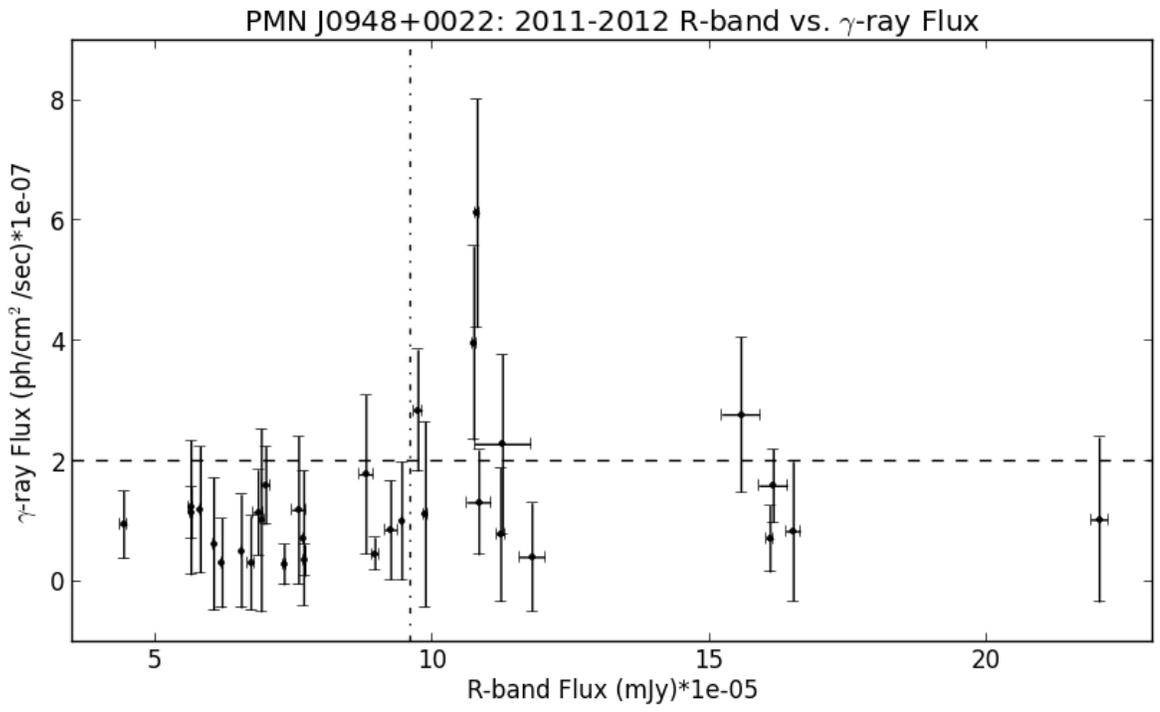}\caption{A plot of concurrent R-band and linearly-interpolated gamma-ray fluxes.
\label{fig:optflux_vs_gammaflux}}
\end{figure*}

The present observations allow us to evaluate if this is the case
for the \textgreek{g}-ray/optical flux variations observed for PMN
J0948+0022. In Fig. \ref{fig:IntGamma_with_opt} we have plotted the
\textgreek{g}-ray and optical light curves for 2011-2013. A possible
correlation between the optical and \textgreek{g}-ray regimes is suggested
in parts of the dual lightcurve. To further evaluate this, in Figure
\ref{fig:optflux_vs_gammaflux} we have plotted the optical flux versus
the \textgreek{g}-ray flux. While we do not see any evidence of a
strong correlation, we do note that above a certain threshold in the
\textgreek{g}-ray brightness (approximately 2{*}10\textsuperscript{-7}
ph cm\textsuperscript{-2} s\textsuperscript{-1}), we consistently
see the object in an elevated optical state (approximately 9.5{*}10\textsuperscript{-5}
mJy or R \ensuremath{\le} 18.7). However, the converse is not true:
we do not always see PMN J0948+0022 in an elevated \textgreek{g}-ray
state when it is optically bright - even during periods of extended
and dramatic brightening, such as that of the period in March 2013
as detailed in Figure \ref{fig:orphan_flares}. The absence of a strong
correlation suggests that there is substantial turbulence present
in the jet and the magnetic field is not highly ordered, in turn suggesting
that no strong shock is present during the time of these observations. 

The picture becomes more complicated if we look at the photopolarimetry
data as displayed in Figure \ref{fig:Rmag_vs_P}. This plot appears
to show a potential bi-modal distribution in the P vs. R-mag plane.
A positive correlation between these values could be seen as supporting
the so-called shock-in-jet interpretation \citep{Marscher et al},
though several data points, especially in the high-magnitude, low-polarization
part of the figure, strongly disagree with this interpretation. However,
as was noted in section 3.1, the two brightest, low-polarization points
occurred during times of rapid flux variation in the object, which
may indicate that these ``errant'' data were the result of turbulence
in part of the jet, rather than a standing shock affecting the entire
optical emitting region.

Our measurements for the doubling timescales in the R-band and \textgreek{g}-rays
- \textgreek{t}\textsubscript{R} and \textgreek{t}\textsubscript{\textgreek{g}},
respectively - provide additional insight into the variability nature
of PMN J0948+0022. As stated in Sections 3.2 \& 3.3, the doubling
timescale in the optical (\textgreek{t}\textsubscript{R}) was 3.99
$\pm$ .15 hours while the corresponding quantity in the \textgreek{g}-ray
regime (\textgreek{t}\textsubscript{\textgreek{g}}) was 3.55 $\pm$
4.29 hours. These values represent a significantly faster doubling
timescale for this object as compared to those presented by \citet{Foschini et al 2012},
who found the values of \textgreek{t} in the optical or \textgreek{g}-ray
bands to be on the order of 2-4 days. The close agreement of the values
presented in this work for \textgreek{t}\textsubscript{R} and \textgreek{t}\textsubscript{\textgreek{g}}
could be used to argue in favor of comparable sizes for the emitting
regions of both the optical and high-energy radiation, implying that
these regions are located close to each other along the jet, though
not necessarily co-spatial. Localized turbulence in part of the jet,
rather than (or in addition to) some sort of standing shock, may better
explain the observed behavior.

It is, perhaps, not surprising that the short variability timescales
of this object have gone undetected in previous studies, as it required
several dedicated optical observing runs and constant monitoring at
high energies to obtain data of sufficient quality to deduce such
values for this work. Over two dozen nights of high-cadence, focused
observations in the optical were required to obtain the two nights
of data that allowed us to calculate \textgreek{t}\textsubscript{R},
whilst an exceptional episode of \textgreek{g}-ray emission in terms
of both flux and confidence in the measurement was required to determine
\textgreek{t}\textsubscript{\textgreek{g}}. Clearly, the observed
behavior of PMN J0948+0022 is very complex and may require the application
of models that take into account turbulence as well as shocks in the
jet to explain this behavior adequately. All of this underscores the
important role that dedicated, long-term monitoring programs can play
in studying objects of this type.

\section{Summary}

In this work we have identified many common properties of blazars
which seem to be present in PMN J0948+0022. These include a strong
linear polarization which is highly variable in both the percentage
of polarization (P) and orientation of the electric vector (EVPA),
an optical flux which varied by more than 2.88 magnitudes over the
twenty seven month observation period, infrared/optical flaring which
was observed to have no \textgreek{g}-ray counterpart, and upper limits
for the doubling timescales in both the optical and \textgreek{g}-ray
regimes which were both measured to be very fast (around 4 hours).

We have also observed what may be evidence for localized turbulence
in the jet, in the form of strong brightening events with no corresponding
increase in the value of P, extremely rapid doubling timescales in
the optical and \textgreek{g}-ray regimes, and strong/rapid flares
in the optical \& infrared which have no \textgreek{g}-ray counterpart.

While the present observations do not allow one to definitively confront
the model suggesting that the \textgreek{g}-ray emission is produced
at a shock, down-stream in the jet, some distance from the SMBH, it
does suggests what is required: It will require quasi-simultaneous
\textgreek{g}-ray and optical/photopolarimetric observations during
a major \textgreek{g}-ray/optical flare consisting of a change in
flux significantly greater than a factor of three, such as was observed
in the present campaign. Under these conditions, one should be able
to determine if there is a significant increase in the polarization
(P) and the expected rapid change in the electric vector position
angle (EVPA) accompanying such an outburst. Therefore we encourage
continued photopolarmetric monitoring of this object in order to investigate
the behavior of the polarization during the next major outburst of
PMN J0948+0022.

\acknowledgements{}

The authors would like to thank Svetlana Jorstad and Paul Smith for
their comments, assistance, and technical expertise. 

\clearpage{}
\begin{table}
\centering{}%
\begin{tabular}{|c|c|c|c||c|c|c|c|}
\hline 
MJD & R-mag (err) & P (err) & EVPA (err) & MJD & R-mag (err) & P (err) & EVPA (err)\tabularnewline
\hline 
\hline 
5599.8 & 19.18 (0.02) & 0.86 (0.50) & -32.08 (11.3) & 6008.7 & 19.48 (0.02) & 1.92 (0.39) & -46.7 (0.3)\tabularnewline
\hline 
5602.8 & 19.18 (0.02) & 2.02 (0.30) & -36.9 (19.2) & 6039.8 & 19.07 (0.02) & 3.89 (0.61) & 28.9 (13.5)\tabularnewline
\hline 
5705.7 & 18.21 (0.02) & 1.35 (1.29) & 93.4 (8.6) & 6040.7 & 18.89 (0.03) & 3.13 (0.36) & 9.0 (1.7)\tabularnewline
\hline 
5706.7 & 17.88 (0.03) & 12.31 (1.21) & 22.6 (9.3) & 6059.7 & 19.19 (0.03) & 0.90 (0.28) & 6.2 (62.7)\tabularnewline
\hline 
5708.7 & 18.90 (0.02) & 4.00 (1.51) & 11.7 (42.0) & 6251.9 & 18.94 (0.02) & 2.29 (0.69) & -44.0 (0.1)\tabularnewline
\hline 
5951.8 & 19.05 (0.03) & 1.66 (0.26) & -50.74 (2.6) & 6253.0 & 18.68 (0.02) & 2.82 (0.58) & 15.2 (3.6)\tabularnewline
\hline 
5982.8 & 18.678 (0.03) & 1.70 (0.48) & 19.4 (13.6) & 6300.0 & 18.06 (0.03) & 1.89 (1.26) & 9.0 (6.5)\tabularnewline
\hline 
5983.8 & 18.82 (0.03) & 5.95 (0.50) & -59.3 (2.0) & 6301.0 & 18.86 (0.02) & 2.62 (0.65) & -64.0 (3.0)\tabularnewline
\hline 
5984.7 & 18.61 (0.03) & 2.45 (0.16) & 19.9 (5.3) & 6302.0 & 19.00 (0.04) & 1.17 (0.28) & 79.7 (27.2)\tabularnewline
\hline 
6007.8 & 18.71 (0.03) & 8.17 (0.74) & 58.4 (13.5) & 6393.8 & 18.78 (0.02) & 1.20 (0.11) & 66.14 (1.4)\tabularnewline
\hline 
\end{tabular}

\caption{Photopolarimetric observations of J0948+0022 obtained between February,
2011 and May, 2012. Columns are: (1) time of the observation in MJD
(JD - 2.45e6), (2) optical R-band magnitude and (error), (3) percent
polarization of target and (error), and (4) EVPA and (error).\label{tab:PhotPol-obs}}
\end{table}
\begin{table}
\centering{}%
\begin{tabular}{|c|c|c|c|}
\hline 
JD & R-mag & R err & \# images\tabularnewline
\hline 
\hline 
2455599.79411 & 19.180 & 0.020 & 1\tabularnewline
\hline 
2455602.79159 & 19.179 & 0.024 & 1\tabularnewline
\hline 
2455624.71679 & 19.382 & 0.003 & 7\tabularnewline
\hline 
2455625.62464 & 19.129 & 0.003 & 16\tabularnewline
\hline 
2455626.34594 & 19.198 & 0.005 & 16\tabularnewline
\hline 
2455627.31249 & 19.004 & 0.003 & 28\tabularnewline
\hline 
2455627.80059 & 18.812 & 0.004 & 11\tabularnewline
\hline 
2455647.70671 & 18.761 & 0.002 & 18\tabularnewline
\hline 
2455648.34777 & 18.761 & 0.001 & 68\tabularnewline
\hline 
2455649.27107 & 18.798 & 0.001 & 84\tabularnewline
\hline 
\end{tabular}\caption{A sample of our optical data, binned at 1-day intervals as described
in the text. Columns are: (1) time of the observations in JD, (2)
optical R-band magnitude, (3) uncertainty in the magnitude, and (4)
the number of observations used to create the binned data point. The
full table of 114 values used in this work can be found in the online
version of this manuscript. \label{tab:1_day_bin_opt}}
\end{table}
\begin{table}
\centering{}%
\begin{tabular}{|c|c|c|c|}
\hline 
MJD & R-mag (err) & \textgreek{D}R (err) & \textgreek{D}J (err)\tabularnewline
\hline 
\hline 
6360.5 & 17.271 (0.016) & -1.373 (0.026) & -1.375 (0.028)\tabularnewline
\hline 
6363.5 & 18.644 (0.021) & 1.504 (0.030) & 1.334 (0.028)\tabularnewline
\hline 
6364.5 & 17.140 (0.021) & -0.880 (0.028) & -0.793 (0.028)\tabularnewline
\hline 
6368.5 & 18.020 (0.018) & N/A & N/A\tabularnewline
\hline 
\end{tabular}\caption{Data corresponding to the strong optical flaring we observed in March,
2013. Columns are: (1) time of the observations in MJD, (2) optical
R-band magnitude, (3) difference in magnitude between the measurement
of the current row and next row, and (4) same as (3), but for the
J-band data. \label{tab:smarts_opt_ir}}
\end{table}
\begin{table}
\centering{}%
\begin{tabular}{|c|c|c||c|c|c|c|}
\hline 
\multicolumn{3}{|c||}{R-band Data} & \multicolumn{4}{c|}{6-hr \textgreek{g}-ray Bins}\tabularnewline
\hline 
MJD & M\textsubscript{R}(err) & \textgreek{t}\textsubscript{R} & MJD & Flux (err) & TS & \textgreek{t}\textsubscript{\textgreek{g}}\tabularnewline
\hline 
\hline 
5652.61 & 18.69 (0.02) & \multirow{2}{*}{4.39 (0.19)} & 6292.15 & 2.78 (1.67) & 14.05 & \multirow{2}{*}{2.81 (4.23)}\tabularnewline
5652.79 & 17.92 (0.02) &  & 6292.40 & 12.21 (5.61) & 15.66 & \tabularnewline
\hline 
6298.88 & 18.13 (0.04) & \multirow{2}{*}{3.60 (0.23)} & 6292.65 & 8.26 (7.49) & 12.03 & \multirow{2}{*}{4.28 (7.47)}\tabularnewline
6299.04 & 18.96 (0.02) &  & 6292.90 & 21.81 (8.723) & 14.91 & \tabularnewline
\hline 
\hline 
\multicolumn{2}{|c|}{Average \textgreek{t}\textsubscript{R}} & 3.99 (0.15) & \multicolumn{3}{c|}{Average \textgreek{t}\textsubscript{\textgreek{g}}} & 3.55 (4.29)\tabularnewline
\hline 
\end{tabular}\caption{Data used to calculate doubling/halving timescales for optical and
\textgreek{g}-ray data. Columns are (1) time (MJD) of the optical
observation, (2) magnitude (error) in the R-band, (3) timescale \textgreek{t}\textsubscript{R}
(in hours) calculated on from the two adjacent data points in Column
2, (4) mid-point in time of the 6-hour bin from which (5) the photon
flux (error) in ph{*}cm\textsuperscript{-2}{*}s\textsuperscript{-1}was
derived, (6) the TS value of the aforementioned \textgreek{g}-ray
data, and (7) the timescale \textgreek{t}\textsubscript{\textgreek{g}}
calculated from the adjacent \textgreek{g}-ray data. The bottom row
gives the average values and (uncertainties) for \textgreek{t} in
each waveband.\label{tab:dh_timescale_stats}}
\end{table}

\end{document}